\documentclass[%
 reprint,
 amsmath,amssymb,
 aps,
]{revtex4-2}

\usepackage{graphicx}
\usepackage{dcolumn}
\usepackage{bm}
\usepackage{float}
\usepackage{xcolor}
\usepackage{cleveref}
\usepackage{ulem}
\usepackage{comment}

\begin{document}

\preprint{APS/123-QED}

\title{Basset-Boussinesq history force and inertia are relevant \\ for unsteady particle settling dynamics}

\author{Tomek Jaroslawski}
  \email{tomek@stanford.edu}
\author{Beverley McKeon}%

\affiliation{%
 Center for Turbulence Research, Stanford University}%

\author{Divya Jaganathan}
\author{Rama Govindarajan}
\affiliation{International Centre for Theoretical Sciences, Tata Institute of Fundamental Research
}%

\date{\today}

\begin{abstract}

Our experiments on a sphere falling under gravity in Stokes flow show significant history effects. We observe an algebraic, not exponential, relaxation rate to the terminal velocity, validating the solution to the Basset-Boussinesq-Oseen equation. Unlike in steady Stokes theory, our experiments and theory reveal a vortex ring forming around the sphere and drifting away. As the Reynolds number nears unity, the vortex ring lags behind the sphere, departing from Stokesian theory, though the sphere’s algebraic response persists. These findings are critical for particle interactions in the Stokes limit. 

\end{abstract}

\maketitle

Numerous natural phenomena involve particles settling in fluid flows, whose short-time dynamics are crucial for accurately predicting this transient process. Examples include the vertical sinking of surface phytoplankton as marine snow, which influences the carbon cycle \cite{monroy2017modeling, ardekani2017sedimentation,trudnowska2021marine}, the growth and aggregation, into raindrops, of falling cloud droplets  \cite{pinsky1997turbulence} and the settling of volcanic ash \cite{arenal2018new}. Further, in oncological studies, sedimentation dynamics of cells are essential for sorting and isolation \cite{haddadi2018separation}.

Despite growing evidence for the significance of history forces in unsteady particle and bubble dynamics in viscosity-dominated flows, including theoretical findings of slow algebraic rather than exponential relaxation rates of particles to their asymptotic states in different physical contexts \cite{basset1888treatise,Lovalenti_Brady_1993,candelier_radialMig_2004,Langlois_relaxingParticle_2015}, accurate experimental observations are scarce, and none, to our knowledge, quantitatively validate them against theory. In addition, the flow field generated by an accelerating particle in such flows remains poorly understood and has rarely been measured. 

This letter presents experiments on the dynamics of, and flow around, a settling particle in the Stokes regime. We demonstrate the validity of a recent solution \cite{prasath2019accurate} of the Basset-Boussinesq-Oseen equation (BBOE) for a settling particle and derive the form of the surrounding unsteady flow to explain our observations. Additionally, we investigate the influence of fluid's inertial effects, measured by a Reynolds number $Re_p$ based on particle's size and terminal velocity, on both particle dynamics and the evolving flow field, and examine the applicability of the unsteady Stokes model as inertia increases.
 
The simplest model for an accelerating particle at $Re_p \ll 1$, which accounts for finite-time response of the fluid to the particle's motion, is based on unsteady Stokes theory. It presumes a one-way interaction, where particle dynamics are dictated by the flow without perturbing it, or being influenced by other particles, i.e., the suspension is dilute. For a spherical particle, the dynamics then adheres to the Maxey–Riley equation \cite{maxey1983equation,gatignol1983faxen,auton1988force}. In the particular scenario of sedimentation under gravity in a quiescent flow field, the Maxey-Riley equation reduces to the BBOE, see e.g. \cite{clift2005bubbles}. Vectors $\mathbf{y}$ and $\mathbf{v}$ denote the particle's position and velocity, respectively, and evolve from their initial state $[\mathbf{y}(0), \mathbf{v}(0)]$ according to
\begin{equation}
\dot{\mathbf{y}}={\mathbf{v(t)}},
\end{equation}
and the non-dimensional BBOE
\begin{equation}\label{BBO}
\begin{split}
   R\dot{\mathbf{v}}=-\frac{1}{S}\mathbf{v}(t)+(R-1)\mathbf{g} -\sqrt{\frac{3}{\pi S}}\Biggl\{ \frac{\mathbf{v}(0)}{\sqrt{t}} + \int_0^t \frac{\dot{\mathbf{v}}(s)}{\sqrt{t-s}}ds \Biggl\} ~,
\end{split}
\end{equation}
where $R = (2\beta+1)/3$, is the effective density ratio, accounting for added-mass effects, $\beta = \rho_p/\rho_f$, with $\rho_{p}$ and $\rho_{f}$ being particle and fluid density, respectively, and ${\bf{g}}$ is the gravitational acceleration. The Stokes number is $S =a^2/(3\nu T)$, $a$ being the particle radius, $\nu$ the fluid's kinematic viscosity and $T=\nu/U_{T}^2$ the Oseen time scale, beyond which convective inertial effects of the fluid become significant \cite{sano1981unsteady,mei1992,Lovalenti_Brady_1993}. The particle's terminal velocity, $U_T$, is given by the balance between gravity and the quasi-steady hydrodynamic drag force on the sphere, $U_T = 2(\beta-1)a^2g/9\nu$. The overdots represent Lagrangian derivatives in time ($t$). Equation 2 delineates the motion of a spherical particle in the Stokesian regime. The terms on the right-hand side of \cref{BBO} respectively are the quasi-steady Stokes drag, the gravitational force, and the nonlocal Basset-Boussinesq history (BBH) force. The last is an integral force along the particle's trajectory over its lifetime, stemming from differences in acceleration between the particle and the surrounding fluid. Boussinesq \cite{boussinesq1885resistance} first introduced the corrective term to Stokes' approximation for non-uniform flow, which Basset \cite{basset1888treatise} formalized three years later. Oseen \cite{oseen1927neuere} subsequently added further corrections.

Analytical and numerical solutions of BBOE are seldom pursued due to the inherent complexity of handling nonlocality in such an integro-differential equation. Incorporating the force in numerical solutions of general particulate flows entails storage requirements that escalate rapidly over time. Consequently, the significance of the BBH force in particle dynamics has remained relatively under-explored. In recent theoretical progress, Prasath et al. \cite{prasath2019accurate} showed that the BBOE and more generally the Maxey-Riley equation, grounded in unsteady Stokes theory, can be favorably transformed into a local description: a one-dimensional heat equation with non-trivial boundary conditions. The BBOE is formally valid for \(t/T <\sim 1\), when fluid's convective inertial effects are negligible [$O(Re_p)$]. The solution to BBOE for a general initial condition is given in \cite{prasath2019accurate}.
For the special case of zero initial conditions, which is the case here, the solution to the BBOE was provided earlier in the form of sums of error functions \cite{clift2005bubbles}.

One significant implication of the unsteady theory is that a particle settling under gravity approaches its terminal velocity at a rate proportional to \(t^{-1/2}\) \cite{prasath2019accurate}. This contrasts with the expectation of exponential fast relaxation when the BBH force is disregarded. This finding, along with the vortical structure we find, carry broad implications, relevant to many physical phenomena. For example, the sedimentation and coalescence rates of carbonaceous material and microplastics in the ocean could be significantly overestimated upon neglecting history effects. 

There is a growing body of numerical and experimental evidence that highlights the significance of the BBH force. Particles dropped into vortex cores \cite{candelier2004effect}, microbubbles trapped in standing waves and propelled by ultrasound \cite{toegel2006viscosity,garbin2009history}, and particle dynamics in chaotic advection \cite{daitche2011memory} have all demonstrated susceptibility to history forces. For particles settling under gravity, theoretical works such as Guseva et al. \cite{guseva2016history} have identified the BBH force's impact on the sedimentation of nearly neutrally-buoyant particles ($\rho_p \sim \rho_f)$ in turbulent flows. The presence of the history force leads to distinct trajectories with a slow convergence to an asymptotic settling velocity. At $Re_p > 1$ too, Mordant and Pinton \cite{mordant2000velocity} observed an increased settling time, through experiments and simulations. They found in their simulations that the memory term varies with \(t^{-1/2}\) followed by an exponential decrease. In the Stokes regime, Vodop’yanov et al. \cite{vodop2010unsteady} observed that particles settling under gravity exhibit reduced vertical displacement compared to theoretical predictions in which the BBH force is not taken into account. However, the study neither investigated nor quantified the relaxation rate towards terminal velocity.

\begin{figure}[h]
\includegraphics[scale=0.55]{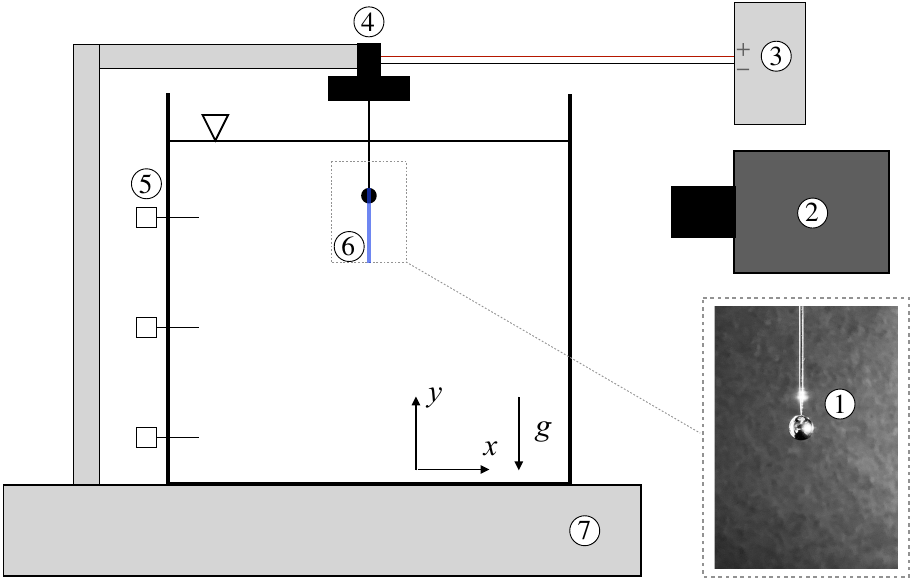}
\caption{Experimental setup:
(1) Particle attached to a magnetized needle. 
(2) High-speed camera.
(3) Power supply.
(4) Electromagnet.
(5) Thermocouples.
(6) PIV laser sheet.
(7) Optical table.}
\label{fig:1}
\end{figure}

\begin{figure}[h]
\includegraphics[scale=0.5]{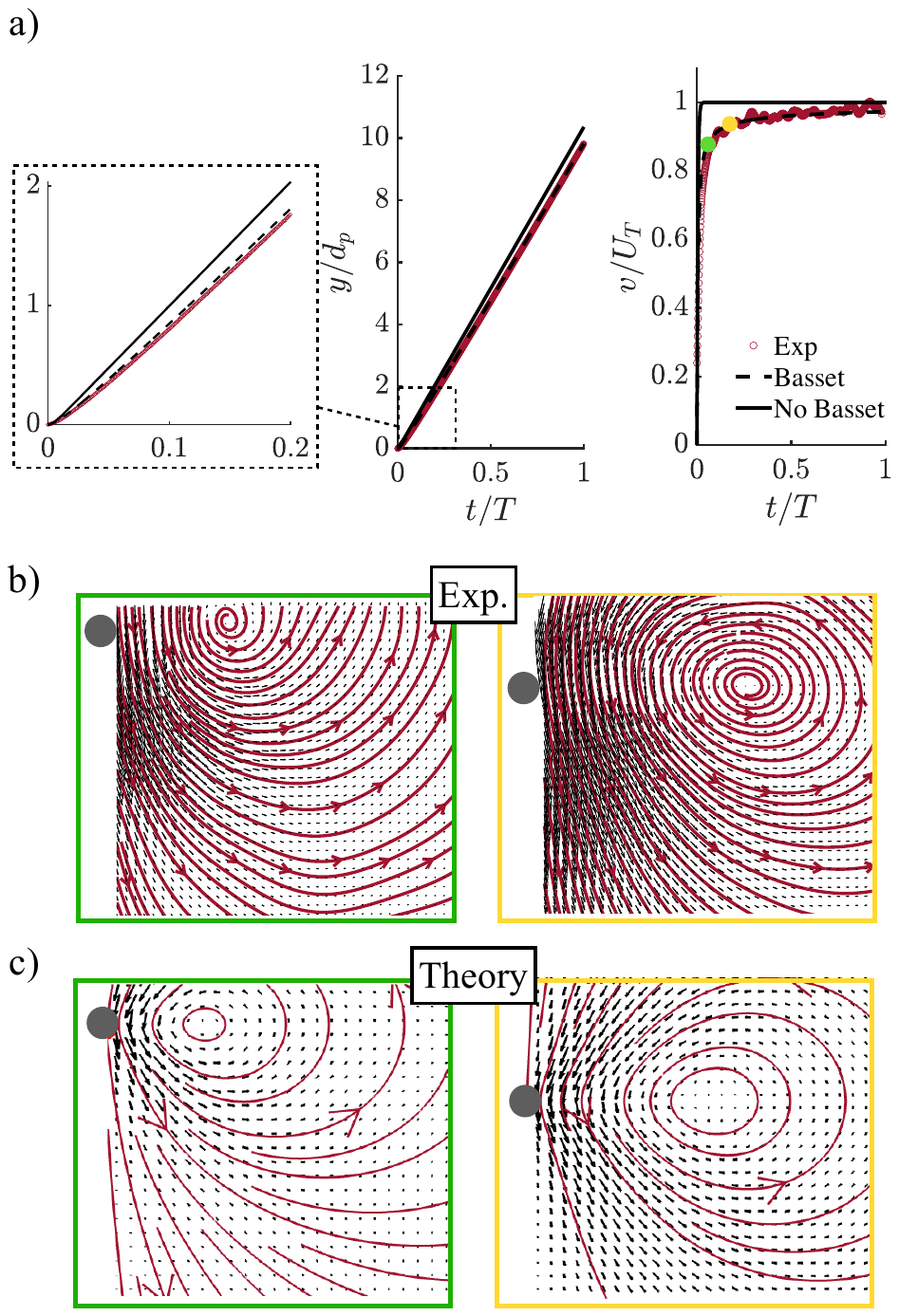}
\caption{
Comparison of experimental data with theoretical predictions for \( Re_{p} = 0.10 \), with terminal velocity \( U_{T} = 0.034 \text{ m/s} \) and Oseen time scale \( T = 0.96 \text{ s} \). (a) Vertical displacement and velocity of the particle over time. Solid and dashed lines: theoretical predictions excluding and including the BBH force, respectively. Experimental data are indicated by markers. (b) Experimental PIV snapshots (cross-section) and (c) theoretical predictions of the unsteady flow field around the particle in the laboratory frame using \cref{unsteadyStreamfunction}, with overlaid streamlines. Green and yellow points in (a) correspond to the snapshots in (b) and (c), respectively.}
\label{fig:2}
\end{figure}

\begin{figure*}
\includegraphics[scale=0.55]{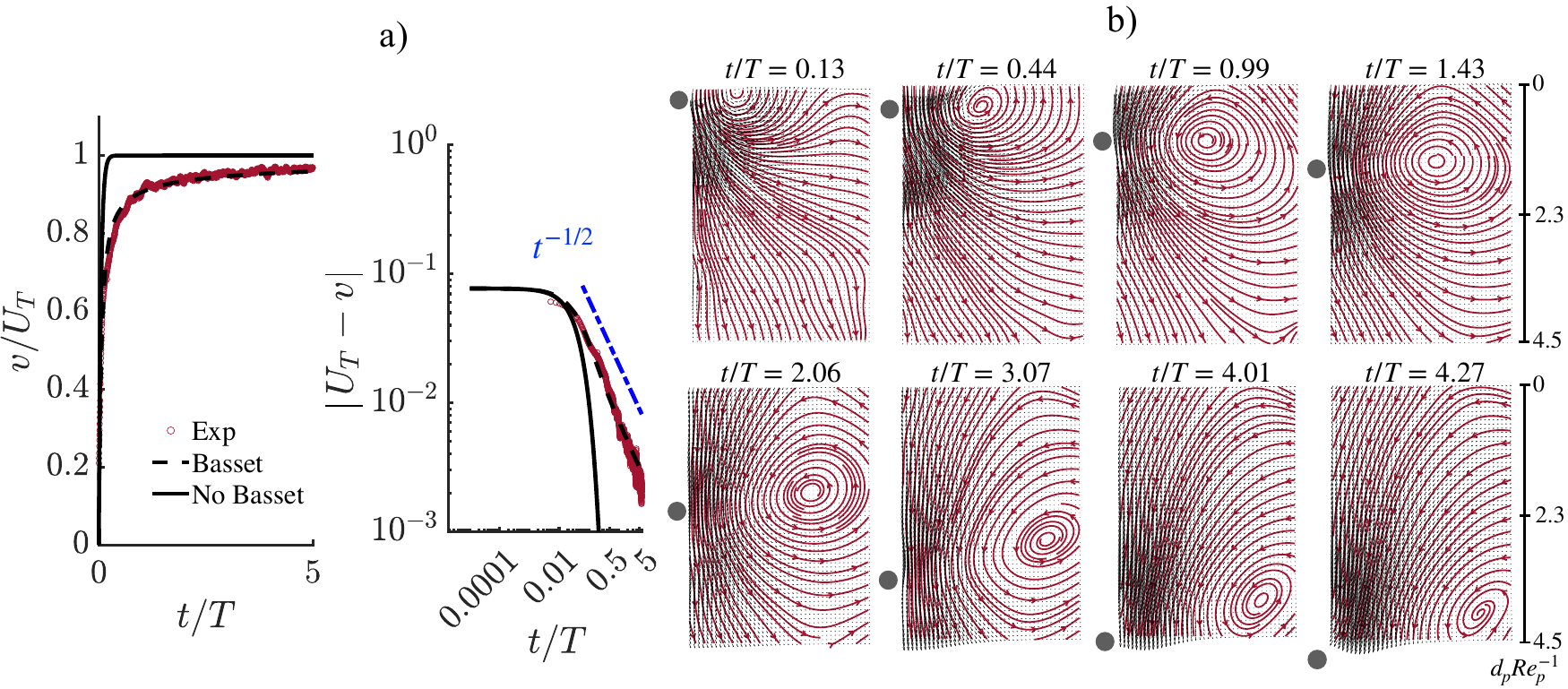}
\caption{Results for \( Re_{p} = 0.32 \), with terminal velocity \( U_{T} = 0.08 \text{ m/s} \) and Oseen time scale \( T = 0.19 \text{ s} \). (a) Vertical velocity versus time and a log-log plot of residual vertical velocity against time. The blue dot-dashed line indicates the theoretical algebraic relaxation rate of \( t^{-1/2} \). Solid lines represent theoretical predictions without the BBH force, while dashed lines include it. Experimental data are shown with markers. (b) PIV snapshots of the unsteady flow field around the particle, with streamlines overlaid for short and intermediate times.}
\label{fig:3}
\end{figure*}

{\textit{Experiments.}}  A series of controlled experiments maintaining the Stokesian conditions were conducted. They specifically targeted the sedimentation of particles from an initial quiescent state ($\mathbf{v}(0)=0$ and $\mathbf{u}(0)=0$). The schematic representation of the experimental setup is presented in \cref{fig:1}. The experiments employed hardened 440C stainless steel particles, with a density of $\rho_{p}=8075 \, \text{kg/m}^{3}$, submerged within a medium of 1000 cSt silicone oil characterized by a density of $\rho_{f} = 997 \, \text{kg/m}^{3}$ (herein considered nominal values, with adjustments made based on temperature monitored throughout the experimental duration). Various particle diameters and densities were tested, and here we present the results of two configurations: \(d_{p} = 3.18\) mm and \(4.76\) mm, which resulted in calculated Reynolds numbers, \(Re_{p} = \frac{U_{T}d_{p}}{v}\), of \(0.1\) and \(0.32\), respectively. In both cases, the particle-to-fluid density ratio is $\beta$ = 8.1. Additional experiments were also conducted for Reynolds numbers of $Re_{p} = 0.76$ (which was outside the Stokes regime for a significant portion of the sedimentation process) with $\beta=8.1$, and $Re_{p}=0.22$ with $\beta=1.6$, achieved through modifications in particle diameter and material composition (see supplemental material). The particle was suspended in the fluid medium utilizing a magnetized tether, with a sting-to-sphere diameter ratio of 0.1 and $0.15d_{p}$ (for $d_{p}$ = 3.18 and 4.76 mm, respectively) and positioned at a depth of 20 and $30d_{p}$ (for $d_{p}$ = 4.76 and 3.18 mm, respectively)  below the free surface to reduce any potential influence from surface and sting-induced effects. A controlled reduction in the magnetic force was systematically applied to facilitate particle release, and it was validated that this process resulted in no rotational or spanwise motions of the released particle. Measures were taken to mitigate any influence stemming from boundary effects by enclosing the experimental domain within a voluminous cubic container, with each dimension being 94 and $140d_{p}$ (for $d_{p}$ = 4.76 and 3.18 mm, respectively), coupled with continuous temperature monitoring of the silicone oil to ensure uniform thermal conditions throughout the experiment. The statistical error and repeatability of the experiment were quantified through reproducing the experiments 20 times. The mean velocity error was found to be 3\% in the early stages where $t/T < 0.1$, and dropped below 1\% for the remainder of the sedimentation process. The uncertainty in the sphere diameter was ±0.03$d_{p}$, while the temperature measurement uncertainty was ±0.5°C, resulting in an uncertainty in the silicone oil viscosity of ±12 cSt. This uncertainty translates to a variation in the mean velocity derived from theory of less than 1\%. Experimental noise was mitigated by smoothing raw trajectories with a Gaussian kernel of width $0.037T$, acting as a low-pass filter with a 200 Hz cut-off frequency.

We measured the trajectory of the particle utilizing a Phantom v2012 high-speed camera, capturing images at a frequency of 2 kHz. Subsequently, we applied a binarization algorithm, leveraging the particle's centroid for trajectory determination. To gain insights into the flow generated, Particle Image Velocimetry (PIV) was employed to quantify the induced flow field. The fluid medium was seeded with 10$\mu$m glass spheres. Utilizing a non-pulsed 450 nm blue laser, the light beam was directed through a series of optics to produce a thin light sheet with a thickness of less than 1 mm oriented in the $x-y$ plane and aligned along the centerline of the sphere. Velocity fields were computed using standard cross-correlation algorithms, with interrogation windows sized at 32 × 32 pixels in the first image and corresponding search windows of 64 × 64 pixels in the second image within each image pair. A 50\% overlap was implemented to achieve the nominal spatial resolution of 16 × 16 pixels or 0.2 × 0.2 $d_{p}$. 

\textit{Theoretical unsteady streamfunction in Stokesian regime.} We posit that the flow field generated by the sphere during sedimentation, $({\bf{u}}, p)$, is governed by the unsteady Stokes equations, $\partial {\bf{u}}/\partial t = -\nabla p + \nabla^2 {\bf{u}} + \sigma (\vec{g}/g)$, which is exact in the vanishing particle Reynolds number limit $Re_p = 0$. Variables have been nondimensionalized by the particle length scale $d_p/2$, and the viscous time scale $d_p^2/4\nu$. For convenience of interpretation, we choose $U_T$ as the velocity scale. Consequently, $\sigma = 9/[2(\beta-1)]$ is the scaled gravitational force density. We seek an azimuthally symmetric solution which satisfies a zero initial condition, ${\bf{u}}(r,\theta,t=0) = {\bf{0}}$, the no-slip boundary condition at the sphere, ${\bf{u}}(r=d_p/2,\theta,t) = {\bf{v}}(t)$, and decay at far-field, ${\bf{u}}(r\rightarrow \infty,\theta,t) = {\bf{0}}$, where the variable $r$ denotes the radial distance measured from the sphere's instantaneous center and $\theta$ is the polar angle. We derive the following exact expression for the scalar non-dimensional streamfunction (scaled by $d_p^2 U_T/4$) under these dynamics, in the laboratory-fixed frame (see supplemental material), 
\begin{equation}\label{unsteadyStreamfunction}
    \psi(r,\theta,t) = \sin^2(\theta)\Big[ \frac{1}{2r}v(t) + \frac{3}{2r} \int_0^t v(\tau)K(t-\tau;r) \:d\tau \Big]~,
\end{equation}
where the kernel $K(\cdot;\cdot)$ is given by
\begin{equation}\label{unsteadyflowKernel}
    K(z;c) = \frac{1-ce^{-\frac{(c-1)^2}{4z}}}{\sqrt{\pi z}}  +\text{erf}\Big\{\frac{(c-1)}{2\sqrt{z}}\Big\},~ c> 1~,
\end{equation}
for non-negative scalars $z,c$. Here, $\text{erf}(\cdot)$ denotes the error function. The above expression generalizes those of \cite{Bentwich_Miloh_1978,Pozrikidis_1989} to a sphere with time-dependent velocity. The time integral in \cref{unsteadyStreamfunction} is the effect of history: of decaying flow perturbations generated by the sphere along its trajectory on the present state. As $t\rightarrow \infty$, the well-known streamfunction from steady Stokes theory can be recovered. The particle velocity $v(t)$ is obtained by solving the BBOE \cref{BBO} exactly following \cite{prasath2019accurate} (see supplemental information), and thereof the streamfunction is obtained from \cref{unsteadyStreamfunction}.

{\textit{Results.}} \Cref{fig:2}a displays the trajectory tracking results, namely the non-dimensional vertical displacement $y/d_{p}$ of the particle at \(Re_{p} = 0.1\) over non-dimensional time $t/T$. The figure compares theoretical predictions with and without the BBH force against experimental data. We observe that incorporating the BBH force yields predictions with reduced displacement, aligning closely with experimental observations, particularly during the early stages of sedimentation where the BBH force holds the most importance. \Cref{fig:2}b depicts the non-dimensional vertical velocity \(v/U_{T}\) as a function of $t/T$, where the experimental data is derived through temporal differentiation (\(dy/dt\)). In the experimental $v/U_{T}$, a clear and distinctive algebraic relaxation rate is evident, which aligns closely with theoretical predictions when the BBH force is considered, in stark contrast to the exponential relaxation observed in its absence. Furthermore, fitting all 20 experimental velocity datasets with a power-law model yields an exponent of $-0.495 \pm 0.095$ for a 95\% confidence interval, thus reinforcing quantitative agreement with  the theoretical prediction of $t^{-1/2}$. Agreement between theoretical predictions that include the BBH force and experimental observations was also established for lower $\beta$ values (see supplemental material). This agreement manifests in both vertical displacement and velocity, as well as in the exponent of the relaxation rate, which remains consistent at $t^{-1/2}$. The PIV velocity vectors, accompanied by plotted streamlines, are shown in \cref{fig:2}b, spanning early stages ($t/T<~1$) of the sedimentation process. Initially, a counterclockwise-rotating vortex emerges close to the particle, which represents a cross-sectional view of a 3D vortex ring. As sedimentation progresses, this vortex gradually shifts away from the sphere along the $x-$direction. The \(x\)-displacement results from the unsteady nature of the initial process and the diffusive flow dynamics around the particle, attributed to the growth of a laminar boundary layer as the particle sediments. In \cref{fig:2}c, snapshots of the velocity vectors with streamlines plotted over different times during early sedimentation, based on the theoretical model presented in \cref{unsteadyStreamfunction}, are shown. The unsteady model successfully captures the vortex formation, handedness, and its drift in the $x$-direction over time, unlike the steady Stokes model, which fails to predict vortex formation.

In this configuration (\(d_{p} = 3.18 \, \text{mm}\) and \(Re_p = 0.10\)), there is a distinct separation between the viscous and inertial time scales: \(a^2/\nu \sim 10^{-3}\) and \(\nu/U_T^2 \sim 1\) seconds, which differ by nearly three orders of magnitude. This allows for a clear delineation between the `Stokes' and `Oseen' regimes. The particle reaches terminal velocity in approximately 0.04 seconds without the BBH force, whereas accounting for the BBH force extends the time to around 2 seconds to reach 90\% of terminal velocity. Additionally, the particle attains a steady state, achieving up to 90\% of the terminal velocity within the Stokesian regime, before the onset of convective effects.  However, for higher \(Re_p\), the theory demonstrates inadequacies in explaining the experimental results.

At higher Reynolds number, the  Oseen time scale \( T \) is smaller, resulting in a reduced physical time interval where \( t/T < 1 \). \Cref{fig:3}a illustrates the vertical velocity of a settling particle over time, demonstrating agreement between theoretical predictions and experimental data at all times. The experimental results show that the BBOE remains applicable up to \(t/T=5\), i.e., beyond its formal range of validity of \(0<t/T < 1\). The theory effectively models the algebraic relaxation rate observed in experiments during this later time. The flow velocity vectors, accompanied by plotted streamlines, are shown in \cref{fig:3}b. Similar to the \(Re=0.10\) case, the vortex core's  vertical position closely follows the particle’s displacement, with an increasingly apparent $x-$direction displacement over time. Around \(t/T \approx 1\), the rate of $x-$direction displacement diminishes, coinciding with the onset of a vertical spatial lag between the vortex core and the center of the particle. At this time, the vortex core has also drifted beyond the Oseen length scale, \( d_{p}Re_{p}^{-1} > 1 \). Another noteworthy observation is the occurrence of asymmetry in the vortex structure around \(t/T \approx 2\), characterized by outward tilting. This behavior could be attributed to the increased influence of inertial effects and induced asymmetry in the vortex structure. However, the low vortex strength at $t>T$ makes its effect on particle motion weak, as confirmed by the agreement between experiment and theory at later times in particle motion. We hypothesize that, over extended periods, the vortex will continue to move away from the sphere, and the streamlines will resemble those of a steady Oseen flow solution, exhibiting fore-aft asymmetry.

\begin{figure}
\includegraphics[scale=0.62]{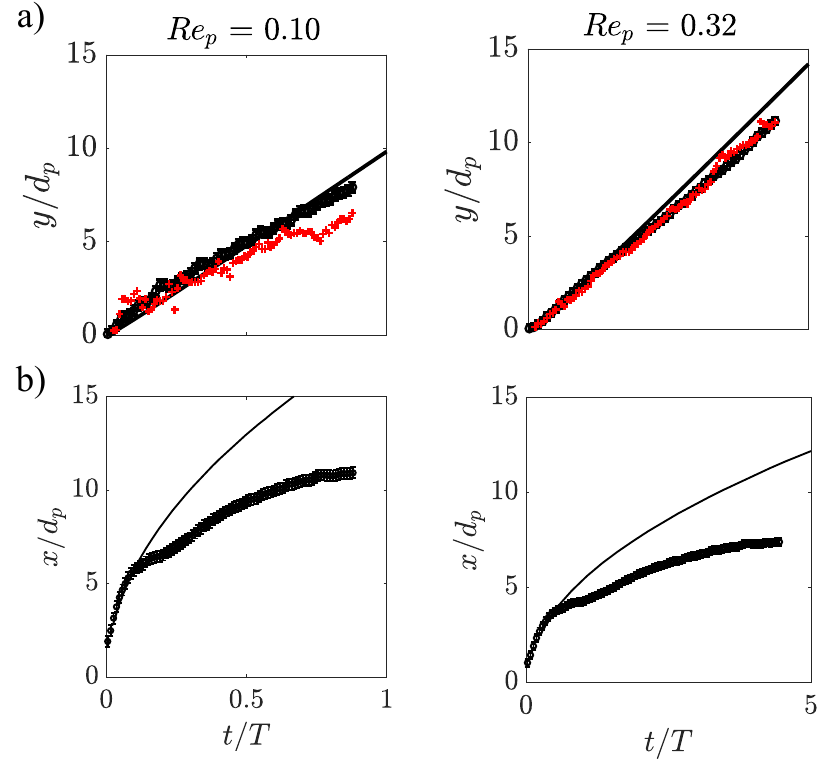}
\caption{Experimental (a) vertical, and (b) horizontal trajectory of the vortex centre (black circles) compared with theoretical predictions (solid black line). Red crosses in (a): $y$-location of the maximum absolute value of \( \mathbf{v}(t) \cdot \nabla \mathbf{u} \). Error bars indicate pixel resolution in PIV data.} 
\label{fig:4}
\end{figure}

In \cref{fig:4}a, we show the vertical trajectories of the particle vortex core from both experiment and theory. Additionally, as an estimate for the convective inertial term \(\mathbf{u} \cdot \nabla \mathbf{u}\), we show the vertical location of the maximum absolute value of \(\mathbf{v}(t) \cdot \nabla \mathbf{u}\) in the field of view, where \(\mathbf{v}(t)\) represents the particle's instantaneous velocity. For \(Re_{p} = 0.10\), the vortex core exhibits horizontal alignment with the particle, as predicted by our unsteady Stokes model in \cref{unsteadyStreamfunction}. However, as \(Re_{p}\) increases to $0.32$, the vortex core, beyond \(t/T \approx 2\), lags behind the particle, a behavior not predicted by our theory, in which fore-aft flow-symmetry is ensured. For both cases, the vertical location of the maximum in the convective term in the flow field follows the vertical location of the vortex, diverging from the particle's path for \( Re_{p} = 0.32 \). The \(x\)-displacement of the maximum convection location was found to be negligible, remaining closely aligned with the particle. To investigate the link with increased inertial effects, we tested a particle with \(Re_{p} = 0.76\), which quickly transitioned out of the Stokes flow regime. We observed the same asymmetry in the vortex dynamics (not shown), attributed to inertial effects. We also evaluated the relative significance of spatially averaged inertial to viscous forces, \( \langle \mathbf{v}(t) \cdot \nabla \mathbf{u} / (\nu \nabla^{2} \mathbf{u}) \rangle_{x,y} \), over time. This analysis revealed an increased relative importance of convective forces compared to viscous ones with larger \(Re_{p}\). In both cases, the vertical location of the maximum convective term correlates with the vertical displacement of the vortex, diverging from the particle's path at \( Re_{p} = 0.32 \). This highlights a link between the vortex core's position and the point of maximum convection in the flow field.  In \cref{fig:4}b, we compare the experimental $x-$displacement of the vortex with theoretical predictions. For both \(Re_{p} = 0.10\) and \(Re_{p} = 0.32\), the theory accurately models experimental observations at early times but soon afterwards, a divergence between experiment and theory occurs. We do not know the reason for this difference.

We show that the unsteady flow structures generated by settling particles in Stokesian conditions exhibits a pronounced sensitivity to inertial effects, which correlates inertial forces with the asymmetries in the particle-induced vortex. The settling dynamics of the particles investigated here are consistent with theoretical predictions, but critically, our results indicate that the BBOE’s applicability to predict the particle's motion extends beyond its formal validity, specifically beyond the Oseen time scale. Our analytical description of the unsteady flow field captures vortex formation and displacement at short times. The intricate dynamics of these unsteady vortices will be crucial in understanding particle-to-particle interactions, necessitating their consideration in the refinement of models and predictive frameworks for a broad spectrum of particle-laden flows.

\textit{Acknowledgements.} The author(s) would like to thank the Isaac Newton Institute for Mathematical Sciences, Cambridge, for support and hospitality during the program \textit{Mathematical aspects of turbulence: where do we stand?} where this work was initiated. This work was supported by EPSRC grant no EP/R014604/1. RG \& DJ acknowledge support of the Department of Atomic Energy, Government of India, under project no. RTI4001.

\newpage

\bibliography{apssamp}

\end{document}